\documentstyle[aps,twocolumn]{revtex}
\input epsf
\input psfig
\begin{document}
\draft
\title{Enhanced absorption Hanle effect on the $F_{\rm g}=F
\to F_{\rm e}=F+1$ closed transitions}

\author{F. Renzoni$^{(1)}$, C. Zimmermann$^{(2)}$,
        P. Verkerk$^{(3)}$, and E. Arimondo$^{(4)}$}

\address{\ $^{(1)}$ Institut f\"ur Laser-Physik, Universit\"at
Hamburg, Jungiusstrasse 9, D-20355 Hamburg, Germany}

\address{\ $^{(2)}$ Physikalisches Institut, Universit\"at
T\"ubingen, Auf der
Morgenstelle 14, D-72076 T\"ubingen, Germany}

\address{\ $^{(3)}$ Laboratoire de Physique des Lasers, Atomes et Mol\'ecules;
U.F.R. de Physique, Universit\'e des Sciences et Technologie de Lille,
59655 Villeneuve d'Ascq Cedex, France}

\address{\ $^{(4)}$ Istituto Nazionale per la Fisica della Materia
and Dipartimento di Fisica, Universit\`a di Pisa, I-56126 Pisa,
Italia}

%\date{\today{}}

\maketitle

\begin{abstract}
We analyse the Hanle effect on a closed $F_g\to F_e=F_g+1$ transition.
Two configurations are examined, for linear- and circular-polarized
laser radiation, with the applied magnetic field collinear to the
laser light wavevector.
We describe the peculiarities of the Hanle signal for
linearly-polarized laser excitation, characterized by narrow bright
resonances at low laser intensities.
The mechanism behind this effect is identified, and numerical
solutions for the optical Bloch equations are presented for different
transitions.
\end{abstract}

\section{Introduction}

Low frequency coherences, between Zeeman or hyperfine split levels,
play an important role in laser spectroscopy and quantum optics.
Phenomena as coherent population trapping \cite{ari_rev},
electromagnetic-induced transparency\cite{harris_rev}, laser without
inversion\cite{zibrov}, nonlinear susceptibility and refractive index
enhancement \cite{harris_90}, steep dispersion \cite{li},
ultra-low group velocity propagation \cite{hau_99} allow us to
realize a coherent control of the absorptive and dispersive properties
of an atomic vapor.
Zeeman coherences are also a basic ingredient of sub-Doppler \cite{cohen}
and sub-recoil \cite{vscpt} laser cooling mechanisms.
Very recently an experimental and theoretical
investigation of a multilevel atomic system with two optical fields
have revealed a new coherent feature of electromagnetic enhanced
absorption \cite{akulshin}. The combined action of the two coherent
optical fields produce a low frequency Zeeman coherence that increases
the atomic absorption. The subnatural linewidth of the resonance
observed scanning the frequency difference between the two optical fields is
determined by the coherence relaxation rate.
In this work we investigate a closely related enhanced absorption
associated to the low frequency Zeeman coherence in a
degenerate two level system and detected in a Hanle effect configuration.

In the original Hanle effect \cite{corney,strumia}, the resonance
fluorescence of an atomic sample is depolarized by an applied
magnetic field as result of the destruction of the excited-state
Zeeman coherences. In that case the Hanle resonance width is the
natural linewidth of the excited state.
In the case of the Hanle effect in the ground state \cite{cct69}, the
width of the Hanle signal is limited only by the relaxation rate of
the ground state and subnatural resonances can be realized. The
low-frequency coherences result also in the modification of the total
fluorescence intensity, and Hanle resonances of both negative and
positive sign have been observed.

The case of a $F_g\to F_e=F_g, F_g-1$ transition has been extensively
studied \cite{picque,strumia2,mclean,giordano} and more recent work
\cite{smirnov,renzoni} has pointed out the strict connection between
Hanle effect on these transitions and coherent population trapping.
For zero magnetic field the atoms are optically pumped into a
nonabsorbing (or dark) state.  An applied magnetic field destroys
the nonabsorbing state and produces an absorption with a linewidth
determined by the relaxation rate of the created Zeeman coherence.
Thus the Hanle effect in the ground state of an atomic system appears
as a decrease in the atomic absorption, with a minimum centered at
zero magnetic field and a linewidth determined by the relaxation
rate of the ground state coherences. However anomalous Hanle effect
lineshapes may be observed under different  experimental conditions.
For instance, for the  transition $2 \to 1$ in neon the circulation
of Zeeman coherence between excited and ground states produced an
inversion in the sign of the Hanle effect, with an increased absorption
at zero magnetic field \cite{ducloy}.

The study of the Hanle effect on a $F_g\to F_e=F_g+1$ transition
leads to different results  depending whether the considered
transition is open or closed. For the {\it open} $F_g=1\to F_e=2$
transition of the sodium D$_1$ line under resonant laser excitation,
the fluorescence profile is given by a nearly Lorentzian curve
\cite{renzoni}. On the other hand, in the case of the {\it closed}
$F_g=2\to F_e=3$ transition of the sodium D$_2$ line line, narrow
resonances of positive sign have been observed superposed to the
homogeneous broadening Lorentzian profile \cite{brand}.
Similar observation have been reported also by Fischer and Hertel
\cite{fischer}, who explained the effect as a disturbance of the
optical pumping process by the Larmor atomic precession.
Those authors connected the linewidth of the fluorescence
increase to the pumping time, but they did not perform a lineshape
investigation.
Very recently experimental results of an enhanced absorption in
rubidium atoms excited by diode lasers on $F_{\rm g} \to
F_{\rm e}=F_{\rm g}+1$ transitions have been obtained
 by two different groups \cite{alzetta,tubi}.

Narrow enhanced-absorption Hanle effect lineshapes have been
observed also for in molecules\cite{weber},
but the mechanism of their production has not been completely defined.

In the present work we analyse the ground-state Hanle effect on a
closed $F_{\rm g} \to F_{\rm e}=F_{\rm g}+1$ transition, with the
standard configuration of an applied linear-polarized laser radiation
and magnetic field collinear to the laser light wavevector. We describe
the peculiarities of the Hanle signal for a closed atomic transition,
and identify the mechanism responsible for narrow bright resonances
at low laser intensities.
We provide a connection with the laser cooling processes where ground
state coherences play an important role. We derive the lineshape for
enhanced absorption Hanle effect with linewidth determined
by the ground state optical pumping time. We investigate the bright
resonance in the Hanle effect either detected  in the fluorescence
emission from the absorbing atoms, as in the experiment of Ref.
\cite{alzetta}, or detected in the light transmitted through the
absorbing medium. Similar constrasts for the bright resonance
are obtained in the two detection schemes.

The paper is organized as follows. In Section II the theoretical basis of
the Optical Bloch Equations (OBE) is introduced.
Section  III presents an analytical solution of the OBE valid in the limit
of weak applied laser electric field. The analytical approach allow us to
derive an expression for the enhanced absorption lineshape.
Section IV presents numerical solutions valid for different atomic
transitions. Finally, in Section V conclusions are drawn.

\section{Optical Bloch Equations}

We consider atoms illuminated by a linearly polarized laser
light,  resonant with the $F_g \to F_e$ transition
of the D$_2$ line. As in the standard Hanle effect configuration, a
static magnetic field $B$ can be applied collinear to
the laser light propagation, the $Oz$  direction.
The laser electric field propagating is given by
\begin{eqnarray}
\vec{E}(z,t) &=& {{\cal E}\over 2}\vec{\epsilon}_x e^{i(k z - \omega
t)}+c.c.\nonumber \\
& =& {\sqrt{2}{\cal E}\over 4}\left( \vec{\epsilon}_{\sigma^+}+
\vec{\epsilon}_{\sigma^-}\right) e^{i(k z - \omega t)}+ c.c.
\end{eqnarray}
with $\vec{\epsilon}_j$ the unit vector of the $j$ polarization.
For comparison, also the case of circularly polarized laser light
(say $\sigma^+$):
\begin{equation}
\vec{E}(z,t) = {{\cal E}\over 2}\vec{\epsilon}_{\sigma^+}
e^{i(k z - \omega t)}+ c.c.
\end{equation}
will be examined.
In writing the optical Bloch equations
\begin{equation}
\dot{\rho}={1\over i\hbar}\left[ H,\rho \right] + \hat{\cal R}\rho
\end{equation}
we consider only the contribution to the relaxation operator
$\hat{\cal R}$
due to the spontaneous emission (index SE).
By choosing the quantization axis parallel to the magnetic field,
the OBEs have the following form ($|e_j\rangle = |J_e,I,F_,j\rangle,
|g_j\rangle = |J_g,I,F_g,j\rangle$):
\begin{mathletters}
\begin{eqnarray}
\dot{\rho}_{e_i e_j} &=& - [i\omega_{e_i e_j}+
\Gamma_{F_e\to F_g} (1+\alpha_{F_e \to F_g; F_{g^{'}}}) ]
\rho_{e_i e_j}\nonumber \\
&&+{i\over \hbar}\sum_{g_k}\left( \rho_{e_i g_k} V_{g_k
e_j}- V_{e_i g_k}\rho_{g_k e_j} \right) \label{bloch1}\\
\dot{\rho}_{e_i g_j} &=& - \left[ i\omega_{e_i g_j}+
{1\over 2} \Gamma_{F_e\to F_g} (1+\alpha_{F_e \to F_g; F_{g^{'}}})
\right] \rho_{e_i g_j}\nonumber\\
&&+{i\over \hbar}\left( \sum_{e_k}\rho_{e_i e_k} V_{e_k g_j}-
\sum_{g_k}V _{e_i g_k}\rho_{g_k g_j}\!\! \right) \label{bloch2} \\
\dot{\rho}_{g_i g_j} &=& - i\omega_{g_i g_j}
\rho_{g_i g_j} +
{i\over \hbar}\sum_{e_k} \left(  \rho_{g_i e_k} V_{e_k g_j}-V _{g_i
e_k} \rho_{e_k g_j} \right)\nonumber\\
&& + \left( {d\over dt}\rho_{g_i g_j}
\right)_{SE}~.
\label{bloch3}
\end{eqnarray}
\end{mathletters}
The quantities $\omega_{\alpha_i,\beta_j}$, with $\alpha , \beta
=(e,g)$,
represent the frequency separation between the energies of levels
$\alpha_i$ and $\beta_j$,
\begin{equation}
\omega_{\alpha_i,\beta_j} = {E_{\alpha_i}-E_{\beta_j}\over \hbar},
\end{equation}
where the energy for the Zeeman level of the ground or excited state
with quantum number $m_i$, including the Zeeman splitting due to the
applied magnetic field, is given by
\begin{equation}
E_{\alpha_i} = \delta_{\alpha,e} \hbar \omega_{o} +g_{\alpha}\mu_B
m_{i}B.
\end{equation}
Here $g_{\alpha}$ represents the gyromagnetic factor of the ground or
excited state, and $\mu_B$ the Bohr magneton.
$\Gamma$ is the total spontaneous emission rate for any excited level,
$\Gamma_{F_e \to F_g}$  denotes the spontaneous decay rate on the
$F_e \to F_g$ transition and $\alpha_{F_e \to F_g; F_{g^{'}}}$ the
ratio between  the spontaneous decays on the $F_e \rightarrow
F_{g^{'}}$ and
$ F_e \rightarrow F_g $ transitions. This ratio is given by

\begin{equation}
\alpha_{F_e \to F_g; F_{g^{'}}} ={\Gamma_{F_e \rightarrow F_{g^{'}}}
\over
\Gamma_{F_e \rightarrow F_g} } =
{2 F_{g^{'}} + 1\over 2 F_g + 1}{
\left\{ \matrix{
J_g & F_{g^{'}} & I \cr
F_e & J_e & 1 \cr } \right\}^2   \over
\left\{ \matrix{
J_g & F_{g} & I \cr
F_e & J_e & 1 \cr } \right\}^2}.
 \label{alpha}
\end{equation}
$V_{e_k,g_j}$, the matrix element of the atom-laser interaction
Hamiltonian for linearly polarized laser radiation, in the dipole
and rotating wave approximation is given by
\begin{equation}
V_{e_k, g_j} =-{{\cal E}\over 2}\langle ek|\vec{d}\cdot
\vec{\epsilon}_x|gj
\rangle e^{-i\omega t}= -{ {\cal E}\sqrt{2}\over 4} \langle ek|
d_{+1}+d_{-1}|gj\rangle e^{-i\omega t}.
\end{equation}
\\
The matrix elements of the spherical
components $d_q$ ($q=0,\pm1$) of the electric dipole moment can be
written as:
\begin{eqnarray}
&&<J_e,I,F_e, M_e|d_{q}|J_g,I,F_g,M_g> = {\cal D} (-1)^{J_e
+I+F_e+F_g-M_e+1}\cdot \nonumber \\ 
&\cdot& \sqrt{(2F_g+1)(2F_e+1)} \left( \matrix{
F_g & 1 & F_e \cr
-M_g & q & M_e \cr } \right) \cdot
\left\{ \matrix{
F_e & 1 & F_g \cr
J_g & I & J_e \cr } \right\} \label{dipolo}
\end{eqnarray}
where ${\cal D } = (J_g||d||J_e)$ is the reduced dipole moment,
having indicated with round
and curly brackets  the $3j$ and $6j$ symbols respectively.
The Rabi frequency $\Omega_{\rm R}$ associated to the
strongest electric dipole moment for the $F_{g} \to F_{e}=F_{g}+1$
transition is given by
\begin{equation}
\Omega_{\rm R}= \frac{\sqrt{2}}{2}
{ \langle J_e,I,F_e,M_e=F_e|d_1|J_g,I,F_g,M_g=F_g\rangle{\cal E} \over
\hbar}.
\end{equation}
The spontaneous emission repopulation terms for the density matrix
evolution are
\begin{eqnarray}
&&\left( {d\over dt}\rho_{g_k g_{k'}}\right)_{SE}\!\!\!=
 (2 F_e + 1) \Gamma_{F_e\to F_g}
\sum_{(q,q'=-F_e,+F_e),(p=-1,1)}\nonumber \\
&&(-1)^{p-k-q'}
\left( \matrix{
F_g & \!\! 1 &\!\! F_e \cr
-k & \!\! p &\!\! q \cr } \right)  \rho_{ e_q e_{q'}}
\!\left( \matrix{
F_e &\!\! 1 &\!\! F_g \cr
-q' &\!\! -p &\!\! k' \cr } \right).
\end{eqnarray}
To eliminate the fast oscillation with the laser frequency $\omega$ we
introduce the quantities $\tilde{\rho}$ and $\tilde{V}$ defined as:
\begin{mathletters}
\begin{eqnarray}
\tilde{\rho}_{e_i e_j} &=& \rho_{e_i e_j}\\
\tilde{\rho}_{g_i g_j} &=& \rho_{g_i g_j}\\
\tilde{\rho}_{e_i g_j} &=& \rho_{e_i g_j} e^{i \omega t}\\
\tilde{V}_{e_i g_j} &=& V_{e_i g_j} e^{i \omega t}.
\end{eqnarray}
\end{mathletters}
Thus the matrix elements $\tilde{V}_{e_i g_j}$ are  time-independent
and the density-matrix equations in terms of $\tilde{\rho}$ and
$\tilde{V}$ can be easily solved numerically.

\section{Lineshape Qualitative discussion}

\begin{figure}
\setlength{\unitlength}{1in}
\begin{picture}(1.3,1.3)
%\put(0,0){\epsfxsize 3in angle 90 \epsfbox{fig1.epsi}}
\psfig{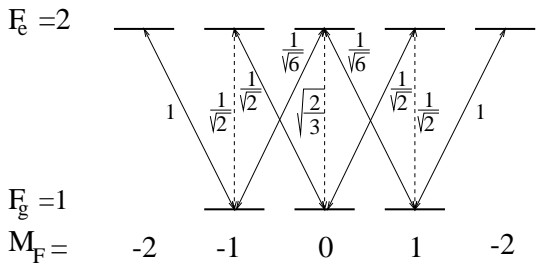}
\end{picture}
\\
\caption{Interaction scheme for a  $F_g =1\to F_e=2$ transition
resonant with linearly polarized laser light for two different choices of the
atomic basis. For a quantization axis collinear to the light wavevector, the
interaction with the light results in $\sigma^{+}, ~ \sigma^{-}$
excitation (solid lines). For a quantization axis collinear to the
light polarization, $\pi$-excitation is produced (dashed lines).
The Clebsch-Gordan coefficients for the different transitions are
reported.}
\label{scheme}
\end{figure}

As a case study to investigate the linewidth of the bright resonance,
we consider an hypothetically closed $F_g=1\to F_e=2$ transition,
with Clebsch-Gordon coefficients as in Fig. \ref{scheme}.
Moreover we will examine the bright resonance by choosing for the
atomic basis two different quantization axis, {\it i.e.} parallel
and perpendicular to the magnetic field direction.

A quick look to the OBE shows that the decay rate for the optical
coherences in Eq. (4b), and of  the excited state populations and
Zeeman coherences in Eq. (4a) is on the order of $\Gamma$, the
natural width. Thus, in order to modify significantly the excited
state populations or coherences related to the excited state, the
magnetic field should be strong enough, {\it i.e.} the Larmor
frequency should be on the order of $\Gamma$. On the other hand,
the natural width does not appear explicitely in Eq. (4c)
for the populations and the coherences in the ground state
multiplicity.
As immediate consequence, the narrow features can be originated from
what occurs in the ground state.
In the case of a closed transition a steady state can be reached and
a particularly simple situation is the limit of weak saturation,
where the laser field can be treated as a perturbation. It is well
known that in this case the excited state can be adiabatically
eliminated, leading to an effective master equation for the density
matrix of the ground state \cite{cohen}. These approximations result
in optical coherences proportional to the amplitude of the laser field
and excited state populations proportional to the light intensity.
The characteristic evolution rate for the ground state populations and
coherences is then given by the parameter $\Gamma^{\prime}$  known as the
optical pumping rate and defined by
\begin{equation}
\Gamma^{\prime} = \frac{\Gamma}{2} \frac{\Omega_{\rm R}^{2}}
{\delta^{2}+\Gamma^{2}/4}.
\label{gp}
\end{equation}
The choice of the quantization axis is also a crucial point. Clearly,
when one introduces a magnetic field, the quantization axis is
usually taken along this field. However, because  the narrow structures
are expected for very weak magnetic fields, a better physical insight
can be obtained in the natural basis given, when B=0, by the light
polarization. In this basis, the steady state populations for the
ground sublevels are $\Pi_{\pm 1}=4/17$ and $\Pi_{0}=9/17$. The atoms
accumulate mostly in the level more interacting  with the light field
(bright state), leading to a maximum in the fluorescence.
When  a magnetic field orthogonal to the quantization axis is
applied, the populations of the ground states will be partially
redistributed, because the Zeeman states are no more eigenstate of
the energy for $B\neq 0$.  Thus the
population of levels weakly coupled to the excited state will increase
at the expenses of the population of the bright states. This redistribution,
leading to a {\it decrease} of the fluorescence rate, explains the bright
resonance.  To couple efficiently the Zeeman sublevels
the Larmor frequency has to be of the order of the relaxation
rate of the coherences or, in the case of a non-zero laser detuning,
of the order of the energy separation due to the light-shift.
Because the relaxation rate of the coherences is $\Gamma^{\prime}$, a
narrow feature is expected, even if a  more careful examination is needed
to give the proportionality coefficient. Notice also the strong
analogy with the Raman spectroscopy of $\sigma^+-\sigma^{-}$ optical
molasses, where a two-photon process couples the Zeeman sublevels
\cite{courtois}.

If the magnetic field is used as quantization axis, a different
description of the process originates with a creation of Zeeman
ground state coherence at zero magnetic field and its destruction
at non zero values of magnetic fields  characterizing the bright
resonance.  In fact the density matrix investigation required to
analyse the Hanle effect with quantization axis along the magnetic
field direction is formally equivalent to that of  the
polarization gradient mechanism  denominated as
$\sigma^{+}-\sigma^{-}$ configuration. In \cite{cohen} it was shown
that the equivalence is realized by replacing the atomic Larmor
frequency in
the applied magnetic field with the Doppler shift of the atoms moving
within the cooling laser field. Thus the equations derived for the
ground-state populations and coherences in those references
eliminating adiabatically the optical coherences and the
excited-state populations and coherences may be used to analyze the
coherence creations in the bright line Hanle effect. Let $C_{\rm r}$
and $C_{\rm i}$ be the real and imaginary parts of the ground state coherence
$\tilde{\rho}_{g_1g_{-1}}$
\begin{equation}
\tilde{\rho}_{g_1g_{-1}}=C_{r}+iC_{i}
\end{equation}
and let $\Pi_{i}$ be  the ground state populations
\begin{equation}
\Pi_{i}=\rho_{g_i g_i}
\end{equation}
For a weak laser intensity, such that the adiabatic eliminations can
be applied, the equivalence with the analysis of ref.\cite{cohen}
allows us to write a closed set of five equations for the ground
state populations and coherences
\begin{mathletters}
\begin{eqnarray}
\dot{\Pi}_{1} &=&
-\frac{5\Gamma^{\prime}}{72}\Pi_1+\frac{9 \Gamma^{\prime}}{72}\Pi_{0}
+\frac{\Gamma^{\prime}}{72}\Pi_{-1}-\frac{\Gamma^{\prime}}{18}C_{r}
-\frac{\delta^{\prime}}{6}C_{i} \label{dens1}\\
\dot{\Pi}_{-1} &=&
\frac{\Gamma^{\prime}}{72}\Pi_1+\frac{9\Gamma^{\prime}}{72}\Pi_{0}
-\frac{5\Gamma^{\prime}}{72}\Pi_{-1}-\frac{\Gamma^{\prime}}{18}C_{r}
+\frac{\delta^{\prime}}{6} C_{i}\label{dens2} \\
\dot{\Pi}_{0} &=& -\dot{\Pi}_{-1} -\dot{\Pi}_{1}
\label{dens3}\\
\dot{C}_{r} &=&
\frac{\Gamma^{\prime}}{24}\Pi_{1}+\frac{\Gamma^{\prime}}{8}\Pi_{0}
+\frac{\Gamma^{\prime}}{24}\Pi_{-1}-\frac{5\Gamma^{\prime}}{12}C_{r}
+\frac{2\mu_gB}{\hbar}C_{i}, \label{dens4}\\
\dot{C}_{i} &=&
\frac{\delta^{\prime}}{12}\left(\Pi_{1}-\Pi_{-1}\right)-
\frac{2\mu_{g}B}{\hbar}C_{r}
-\frac{5\Gamma^{\prime}}{12}C_{i}.
\end{eqnarray}
\label{dens5}
\end{mathletters}
where we have introduced the light shift $\delta^{\prime}$
\begin{equation}
\delta^{\prime} = \frac{\omega - \omega_{o}}{2}
\frac{\Omega_{R}^{2}}{\delta^{2}+\Gamma^{2}/4}
\end{equation}
and $\mu_g = g_g \mu_B$.
Introducing the population difference $d=\Pi_{1}-\Pi_{-1}$ and using
the relation $\Pi_{-1}+\Pi_{1}=1-\Pi_{0}$,  Eqs. (\ref{dens5}) may be
rewritten as
\begin{mathletters}
\begin{eqnarray}
\dot{d} &=& -\frac{\Gamma^{\prime}}{12}d-\frac{\delta^{\prime}}{3}C_{i},
\label{eqr1}\\
\dot{\Pi}_{0} &=&
\frac{\Gamma^{\prime}}{18}-\frac{11\Gamma^{\prime}}{36}\Pi_{0}
+\frac{\Gamma^{\prime}}{9}C_{r},\label{eqr2} \\
\dot{C}_{r} &=&
\frac{\Gamma^{\prime}}{24}+\frac{\Gamma^{\prime}}{12}\Pi_{0}
-\frac{5\Gamma^{\prime}}{12}C_{r}+\frac{2\mu_g B}{\hbar}C_{i}, \label{eqr3}\\
\dot{C}_{i} &=&
\frac{\delta^{\prime}}{12}d-\frac{2\mu_g}{\hbar}BC_{r}
-\frac{5\Gamma^{\prime}}{12}C_{i}.  \label{eqr4}
\end{eqnarray}
\end{mathletters}
whose steady state solution is
\begin{mathletters}
\begin{eqnarray}
C_{r} &=& \frac{5}{34} \left[
1 +\frac{132}{17}\frac{\left(\frac{4\mu_g B}{\hbar}\right)^{2}}
{4\left(\delta^\prime\right)^{2}+5\left(\Gamma^\prime\right)^{2}}\right]^{-1},
\label{res1}\\
\Pi_{0} &=&\frac{2}{11}(1+2C_{r}),\label{res2} \\
C_{i} &=&-
\frac{24\mu_g B}{\hbar}\frac{\Gamma^{\prime}}
{4\left(\delta^\prime\right)^{2}+5\left(\Gamma^\prime\right)^{2}} C_r,
\label{res3}\\
d &=& -4\frac{\delta^{\prime}}{\Gamma^{\prime}}C_{i}.\label{res4}
\end{eqnarray}
\end{mathletters}
The fluorescence signal is proportional to the total excited state
population. In the low saturation limit, the population of each
excited state  may be expressed as a function of the ground state
quantities \cite{cohen}
\begin{mathletters}
\begin{eqnarray}
\rho_{e_{\pm2}e_{\pm2}} &=&
\frac{1}{4}\frac{\Omega_{R}^{2}}{\delta^{2}+\Gamma^{2}/4}
\Pi_{\pm1},\label{exc2} \\
\rho_{e_{\pm1}e_{\pm1}} &=&
\frac{1}{8}\frac{\Omega_{R}^{2}}{\delta^{2}+\Gamma^{2}/4} \Pi_{0},
\label{exc1}\\
\rho_{e0e0} &=&\frac{1}{2}\frac{\Omega_{R}^{2}}{\delta^{2}+\Gamma^{2}/4}
\left[\frac{1}{12}\left(\Pi_{1}+\Pi_{-1}\right)+
\frac{1}{6} C_r
\right].\label{exc3}
\end{eqnarray}
\end{mathletters}
The total excited state population at the steady state $\Pi^{e}_{st}$
is
\begin{equation}
\Pi^{e}_{st}= \sum_{i=-2,2} \rho_{e_ie_i}= \frac{\Omega_{R}^{2}}
{\delta^{2}+\Gamma^{2}/4} \left(
\frac{25}{88}+\frac{3}{44}C_{r} \right).
\label{excpop}
\end{equation}
Thus the variation in the fluorescence with the magnetic field is
associated to the modification of the ground state coherence with
the magnetic field.
The laser interaction produces a large coherence, 5/34, between the
$m_{F}=-1$ and $m_{F}=1$ Zeeman sublevels. Increasing the magnetic
field the coherence $C_{r}$ is reduced from its  maximum value at
$B=0$ to zero. The equation for $C_{r}$
represents a narrow bright resonance whose HWHM $\Delta B$ is
\begin{equation}
\Delta B=\frac{\hbar}{8\mu_{g}}
\sqrt{\frac{17}{33}\left[4\left(\delta^\prime\right)^{2}
+5\left(\Gamma^\prime\right)^{2}\right]}.
\end{equation}
The contrast C, defined as the amplitude of the narrow resonance
divided by the amplitude of the homogenous broadened line, is 3/85,
independent of the Rabi frequency and laser detuning in the limit of
low-saturation. Note however that the Rabi frequency and laser detuning
determine the time scale to reach the steady state, as from the Eq.
(\ref{gp}) for $\Gamma^\prime$.

On the other hand, when the polarization of the light is circular,
the population accumulates in the m=+1 sublevel and a magnetic
field along the quantization axis does not redistribute the
populations. As a consequence, one expects no narrow structures
in the case of a circularly polarized beam with a magnetic field
along the propagation axis.

\section{Numerical solution of the OBE}

To point out out the basic features of the effect of narrow bright
resonances, we have solved the OBE for  a cold atomic sample, where
the laser light excites only one atomic hyperfine transition, with
only the natural broadening. Moreover for a cold sample the interaction
time between laser and atoms may be assumed long enough to consider only
the steady state solution of the density matrix equations.

The bright line has been detected either on the fluorescent light
emitted by the atomic sample as in experiment of Ref. \cite{alzetta},
or on the light transmitted by the atomic sample, as in the experimental
results reported in Ref. \cite{tubi}. For the fluorescence measurements the
bright line appear as an increase of the emitted light. For the transmission
measurements the bright line appears as an additional loss, {\it i.e.}
as a reduced transmission, and an increase in absorption. For a laser
excited closed transition, the atoms reach a steady state, and the
fluorescence intensity $I(B)$ emitted from the cold atomic sample
interacting with the laser beam is proportional to the steady state total
population of the excited state $\Pi^e_{st}$.
Thus the dependence of this quantity on the amplitude of the applied magnetic
field produces the Hanle effect lineshapes observed on the fluorescence
emission.
For the transmission measurements, we characterize the transmission signal
$T$ by considering the imaginary part $\chi^{''}$ of the susceptibility,
{\it i.e.} the loss per unit wavelength.

\subsection{Closed Rb D$_2$ line transitions} \label{mechanism}

\begin{figure}
\setlength{\unitlength}{1in}
\begin{picture}(4,8)
%\put(0,0){\epsfxsize 3in \epsfbox{fig2.eps}}
\psfig{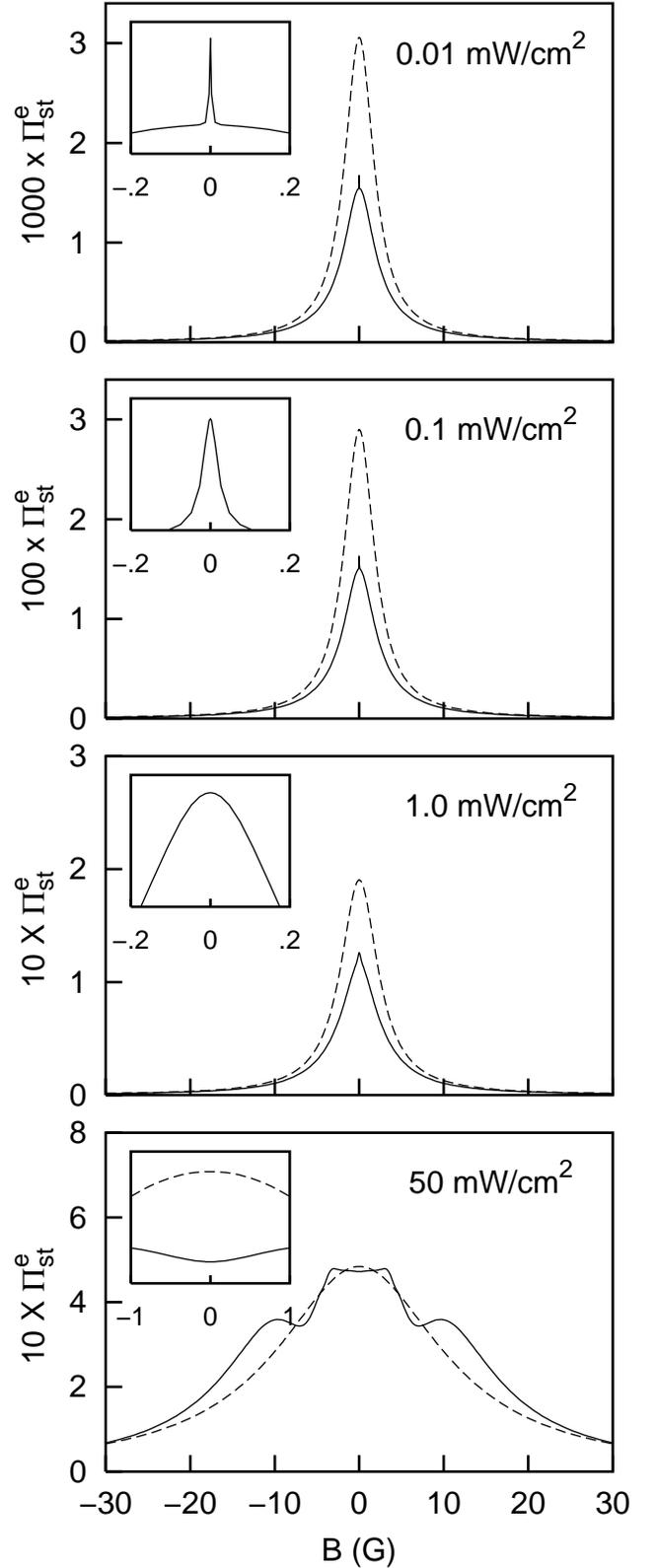}
\end{picture}
\\
\caption{Steady-state total population of the excited state as a
function of the applied magnetic field for different laser intensities.
The calculations refer to the $F_g=2\to F_e=3$ transition of the
$^{87}$Rb D$_2$-line. Solid lines correspond to linearly polarized
laser light, dashed line to circular polarization. The insets evidence
the regions around zero magnetic field.}
\label{num}
\end{figure}

\begin{figure}
\setlength{\unitlength}{1in}
\begin{picture}(4,6)
%\put(0,0){\epsfxsize 3in \epsfbox{fig3.eps}}
\psfig{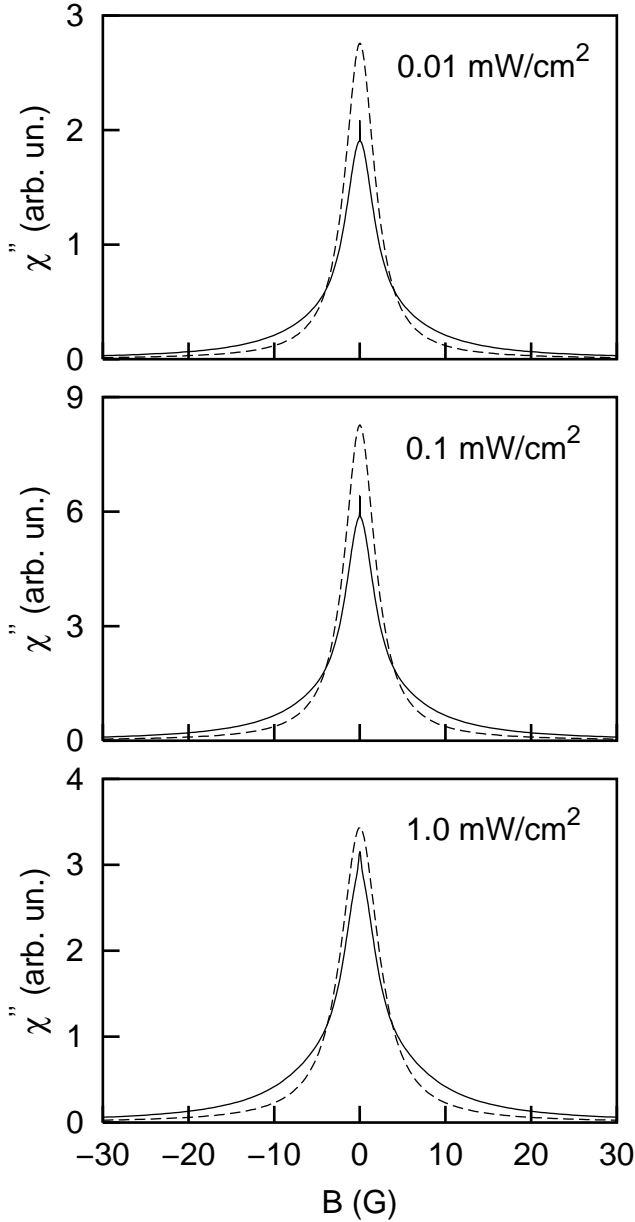}
\end{picture}
\\
\caption{Imaginary part of the total susceptibility versus the magnetic
field $B$ for different laser intensities in the case of the
$F_g=3\to F_e=4$ transition of the $^{85}$Rb D$_2$-line. Solid lines
correspond to linearly polarized laser light, dashed line to circular
polarization.}
\label{transm}
\end{figure}

We have examined the bright line Hanle signals for both the $F_g=2\to
F_e=3$ D$_{2}$ transition of the the $^{87}$Rb isotope and for the
$F_g=3\to F_e=4$ D$_{2}$ transition of the  $^{85}$Rb isotope, with
similar results in the two cases. Results of the numerical calculation
for the total stationary excited-state population $\Pi^{e}_{st}(B)$
for $^{87}$Rb are reported in  Fig. \ref{num} at different laser
intensities.
Results of the numerical calculation for the susceptibility $\chi''$
in the case of  the $F_g=3\to F_e=4$ D$_{2}$ transition of the the
$^{85}$Rb isotope  are reported in Fig. \ref{transm} also for different
laser intensities. In both cases for comparison also the results for
circularly polarized laser light have been reported as dashed lines.
In the case of linearly polarized laser light, for small laser
intensities a narrow bright resonance appears around $B=0$, in both
fluorescence emission and transmission, superposed to a broader line.
That narrow feature is more clearly shown in the insets, and is absent in
the case of  circularly polarized laser field. The contrast $C$ of
the bright resonance is around ten percent for both detection schemes
and is nearly independent of the laser intensity up to intensities of
0.1 mW/cm$^{2}$. At larger laser intensities the contrast decreases.
Furthermore the bright resonance disappears at laser intensities
around few mW/cm$^{2}$.

At high laser intensities, the distribution of the population in the ground
state is irrelevant. It is important instead the number of excited-state levels
which can be effectively populated. For zero magnetic field, the two Zeeman
levels edges of the excited state, with $M_e=\pm 3$, are not populated and
all the allowed transitions $|g,i\rangle \to |e,i\rangle$ are saturated,
producing $\rho_{g_ig_i}\simeq \rho_{e_ie_i}$. Therefore the total
fluorescence intensity $I(0)$ becomes
\begin{eqnarray}
I(0)&=&\Gamma \sum_{i=-2,2} \rho_{e_ie_i} \simeq {\Gamma\over
2}\left(\sum_{i=-2,2}
\rho_{e_i e_i} +\sum_{i=-2,2} \rho_{g_ig_i}\right) \nonumber\\
&=& {\Gamma\over 2}
\end{eqnarray}
independent of the distribution of the atomic population.
For non-zero magnetic field also the two Zeeman edge states can be
occupied as a result of the magnetic-field mixing between excited state
sublevels, so the total excited-state population can increase. This explain
the high-intensity reversed lineshape appearing in Fig. \ref{num}.

\subsection{Open $F_g=1\to F_e=2$ transition of the $^{87}$Rb
D$_2$-line}

\begin{figure}
\setlength{\unitlength}{1in}
\begin{picture}(4,8.1)
%\put(0,0){\epsfxsize 3in \epsfbox{fig4.eps}}
\psfig{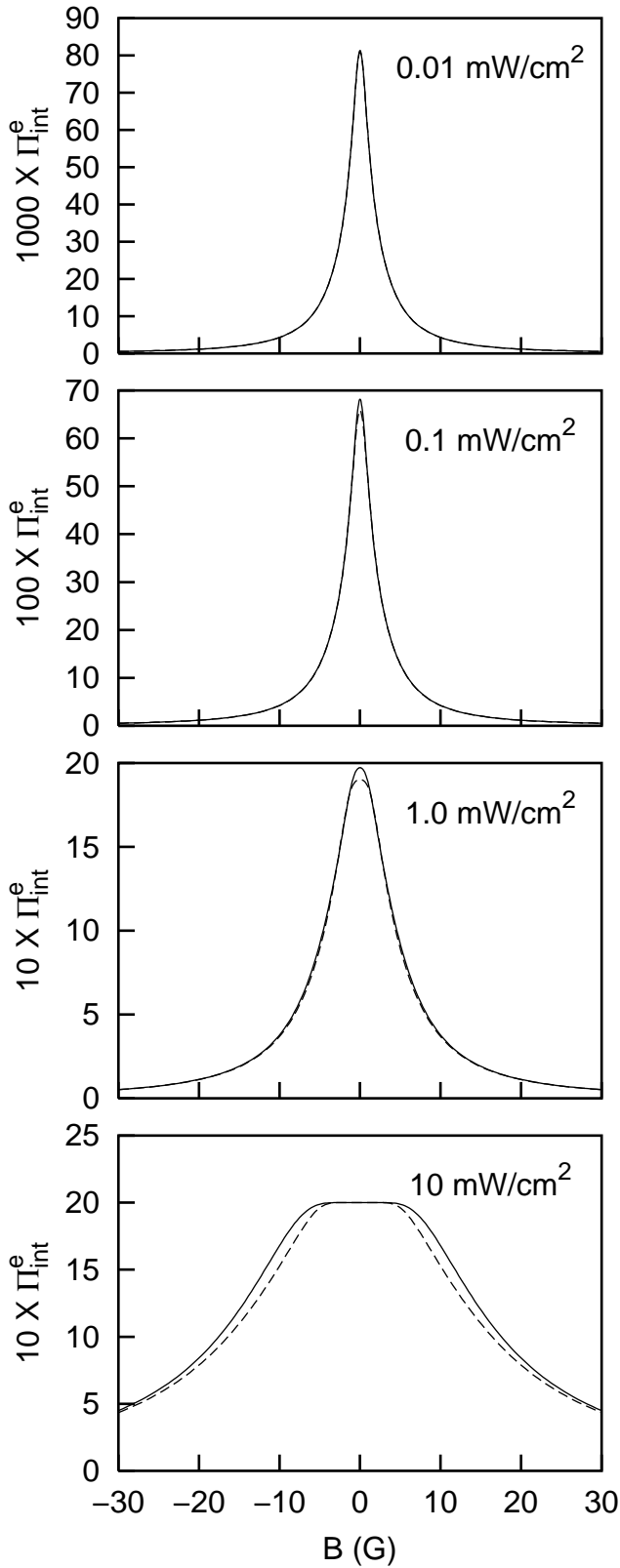}
\end{picture}
\\
\caption{Total population of the excited state integrated over an
interaction time $t_f=100\Gamma$ as a function of the applied magnetic
field for different laser intensities. The calculations refer to the
$F_g=1\to F_e=2$ transition of the $^{87}$Rb D$_2$-line. Solid lines
correspond to linearly polarized laser light, dashed line to circular
polarization.}
\label{num12}
\end{figure}

We consider now the open $F_g=1\to F_e=2$ transition of the $^{87}$Rb
D$_2$-line. For a laser excited open transition, the stationary
excited-state population is zero \cite{renzoni}, and the fluorescence
intensity $I_{\rm int}(B)$ emitted from the atomic sample within the
interaction time $t_{\rm int}$ depends on the time integrated total
population $\Pi^{e}_{\rm int}(B)$ of the excited state defined by
\begin{equation}
I_{\rm int}(B) = \Gamma\Pi^{e}_{\rm int}(B) = \Gamma\int^{t_{\rm
int}}_{0}
\Pi^{e}(B,t) dt.
\label{integral}
\end{equation}
By numerically solving the optical Bloch equation, we calculated the
integrated  excited-state population $\Pi^{e}_{\rm int}$, Eq.
(\ref{integral}), as reported in Fig. \ref{num12}. The plots indicate
that in this case there is no bright resonance. Such a result is in
agreement with that  of Akulshin {\it et al}\cite{akulshin} derived for a
bichromatic configuration: subnatural resonances do not  occurr in open
transitions. In fact, as  discussed previously, the phenomenon of bright
resonances corresponds to the redistribution, via optical pumping, of the
population in the ground state with an accumulation
into the levels maximally coupled to the excited state. The presence
of a
channel of decay out of the levels excited by the laser contrasts this
accumulation: the strong transitions are depleted at a higher rate
than the weak
ones. For a loss rate large enough, the depletion dominates and no
bright resonance occurs. Finally, for  sufficiently large interaction
times and laser intensities, the atomic states are completely depleted
by the laser.

The analysis of Ref. \cite{renzoni} has shown that the time-integrated
excited-state population depends only on the loss parameter
$\alpha_{F_e=2\to F_g=1;F_{g'}}$, and is independent of the magnetic
field, for a range within the homogeneous linewidth. The flat spectra
of Fig. \ref{num12} at large laser intensities are produced by those
dependencies.

\subsection{Closed $F_g=4\to F_e=5$ transition of the Cs D$_2$-line}

In the work of Th\'eobald {\it et al} \cite{giordano} the Hanle
effect on the different hyperfine transitions of the Cs D$_2$-line
was investigated for applied linearly-polarized broadband laser
radiation. Their result for the closed $F_g=4\to F_e=5$ transition is,
at first sight, in disagreement with our general conclusion for a
$F_g=F\to F_e=F+1$ transition.

\begin{figure}
\setlength{\unitlength}{1in}
\begin{picture}(4,6.5)
%\put(0,0){\epsfxsize 3in \epsfbox{fig5.eps}}
\psfig{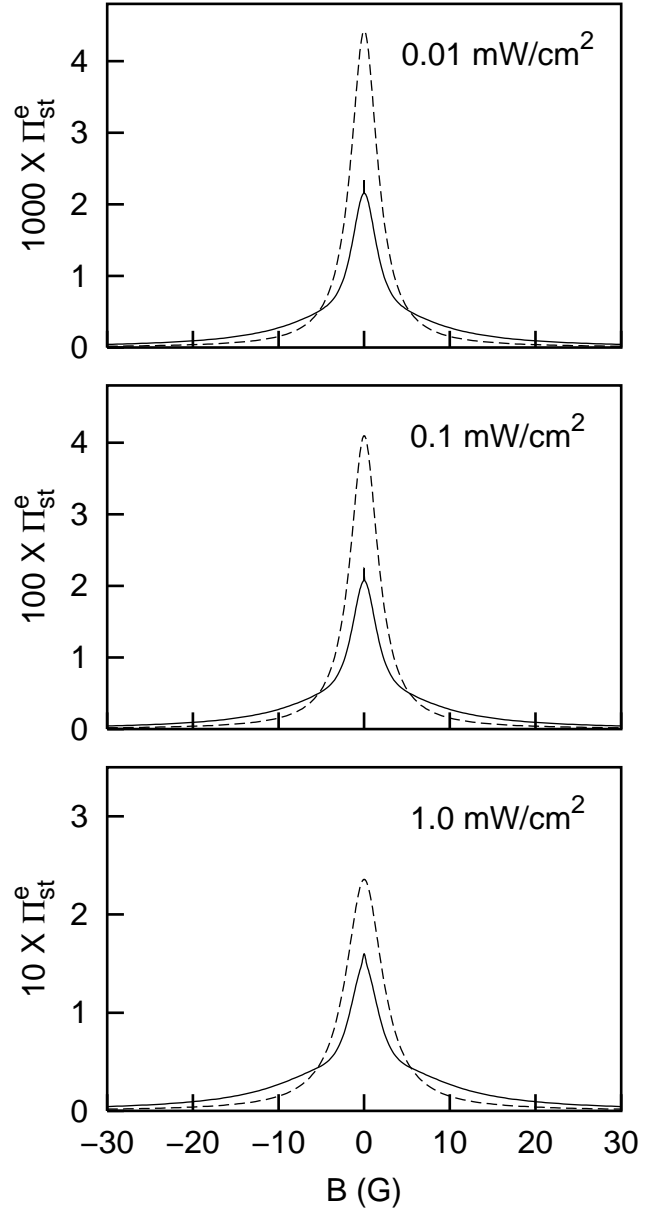}
\end{picture}
\caption{Steady-state total population of the excited state as a
function of the applied magnetic field at different laser intensities,
for the $F_g=4\to F_e=5$ transition of the Cs D$_2$-line. Solid lines
correspond to linearly polarized laser light, dashed line to circular
polarization.}
\label{num45}
\end{figure}

In fact in that experiment the measured
intensity of the fluorescence from the atomic sample was almost constant
around zero magnetic field, for a laser intensity of 2 mW/cm$^2$ and an
interaction time of 12 $\mu$s.
In order to clarify this point, we solved
the optical Bloch equations also for the $F_g=4\to F_e=5$ closed transition
of the cesium D$_2$ line. Results for $\Pi^{e}_{\rm st}$ are shown in Fig.
\ref{num45}.

The dependence of the excited-state population on the magnetic
field at different laser intensities and polarizations is completely
analogous to the behaviour observed for the closed $F_g=F\to F_e=F+1$
transition of the $^{85}$Rb and $^{87}$Rb, therefore confirming the
general validity of this behaviour. However the results
of Fig. \ref{num45} clarify why in the experiment of Th\'eobald {\it
et al}
the fluorescence intensity was almost constant around zero magnetic
field: in the experimental investigation the laser intensity was beyond
the low-saturation regime, in which sharp bright resonances occur.

\section{Conclusions}

We studied theoretically the Hanle effect on closed $F_g=F\to F_e=F+1$
transitions. Both cases of linear- and circular-polarized laser excitation
have been considered.  An analytic solution of the optical
Bloch equations has been given for a closed $F_g=1\to F_e=2$ transition.
Numerical solutions have been obtained for the transitions corresponding
to available experimental data.  

For the examined closed transitions resonant with a linear polarization
laser field, bright resonances, {\it i.e.} sharp increases of the
absorption have been reproduced around zero magnetic field at low laser
intensities.
The enhanced absorption corresponds to the accumulation, via optical
pumping, of the atomic population in the ground states maximally
coupled to the excited state. The application of a magnetic field
perpendicular to the light polarization  redistributes the
population among the ground state sublevels, and results in a decrease
of the fluorescence intensity.
At increasing intensities  of the linear-polarized
laser field, the strength of the  bright resonance is progressively
reduced, and in the saturation regime of the optical transition a
decrease of the absorption around zero magnetic field is produced.

Our calculations for open transitions have confirmed that the bright
resonance does not generally occurs for this kind of transitions. In
effect for open transitions the laser field will empty the
ground state with a rate similar to that needed to redistribute
the ground state population.
Furthermore it should be pointed out that the bright-line production
requires an optical pumping process. Thus, depending on the applied
laser intensity, the interaction time should be long enough to let
the atoms reach their steady state. For instance, the steady state was
not reached in Ref. \cite{brand} where it was demonstrate that for
sufficiently weak light intensities the bright resonance disappears,
because the redistribution of the population in the ground state was not
achieved.

\acknowledgements

We thank S.S. Cartaleva and G. Alzetta for showing us their results
prior to publication. F.R. thanks A. Hemmerich for many useful
discussions and the Deutsche Forschungsgemeinschaft for financial
support under project Li 417/4-1. The {\it  Laboratoire de Physique
des Lasers, Atomes et Mol\'ecules} is a {\it Unit\'e Mixte de Recherche
de l'Universit\'e de Lille 1 et du CNRS}.

\end{document}